\pdfoutput=1
\documentclass[sigconf,10pt]{acmart}
\usepackage{subcaption}
\usepackage{pifont}
\usepackage{multirow}
\usepackage{balance}
\usepackage{flushend}

\newcommand{\red}{\color{red}}
\newcommand{\blue}{\color{blue}}

\newcommand{\sect}{\textsection}

\setcopyright{acmcopyright}
\copyrightyear{2021} 
\acmYear{2021} 
\setcopyright{acmcopyright}
\acmConference[IPSN '21]{The 20th International Conference on Information Processing in Sensor Networks (co-located with CPS-IoT Week 2021)}{May 18--21, 2021}{Nashville, TN, USA}
\acmBooktitle{The 20th International Conference on Information Processing in Sensor Networks (co-located with CPS-IoT Week 2021) (IPSN '21), May 18--21, 2021, Nashville, TN, USA}
\acmPrice{15.00}
\acmDOI{10.1145/3412382.3458255}
\acmISBN{978-1-4503-8098-0/21/05}

\begin{document}

\title[PhyAug: Physics-Directed Data Augmentation for Deep Sensing Model Transfer]{PhyAug: Physics-Directed Data Augmentation for Deep Sensing Model Transfer in Cyber-Physical Systems}

\author{Wenjie Luo}
\affiliation{%
  \department{Singtel Cognitive and AI Lab for Enterprises}
  \institution{Nanyang Technological University}
  \country{Singapore}
}
\authornote{Also with School of Computer Science and Engineering, Nanyang Technological University, Singapore.}

\author{Zhenyu Yan}
\affiliation{%
  \department{Singtel Cognitive and AI Lab for Enterprises}
  \institution{Nanyang Technological University}
  \country{Singapore}
}
\authornotemark[1]

\author{Qun Song}
\affiliation{%
  \department{ERI@N, Interdisciplinary Graduate School}
  \institution{Nanyang Technological University}
  \country{Singapore}
}
\authornotemark[1]

\author{Rui Tan}
\affiliation{%
  \department{Singtel Cognitive and AI Lab for Enterprises}
  \institution{Nanyang Technological University}
  \country{Singapore}
}
\authornotemark[1]


\begin{abstract}
  Run-time domain shifts from training-phase domains are common in sensing systems designed with deep learning. The shifts can be caused by sensor characteristic variations and/or discrepancies between the design-phase model and the actual model of the sensed physical process. To address these issues, existing transfer learning techniques require substantial target-domain data and thus incur high post-deployment overhead. This paper proposes to exploit the first principle governing the domain shift to reduce the demand on target-domain data. Specifically, our proposed approach called PhyAug uses the first principle fitted with few labeled or unlabeled source/target-domain data pairs to transform the existing source-domain training data into augmented data for updating the deep neural networks. In two case studies of keyword spotting and DeepSpeech2-based automatic speech recognition, with 5-second unlabeled data collected from the target microphones, PhyAug recovers the recognition accuracy losses due to microphone characteristic variations by 37\% to 72\%. In a case study of seismic source localization with TDoA fingerprints, by exploiting the first principle of signal propagation in uneven media, PhyAug only requires 3\% to 8\% of labeled TDoA measurements required by the vanilla fingerprinting approach in achieving the same localization accuracy.
\end{abstract}


  
\maketitle

\section{Introduction}
\label{sec:intro}

Recent advances of deep learning have attracted great interest of applying it in various embedded sensing systems. The deep neural networks (DNNs), albeit capable of capturing sophisticated patterns, require significant amounts of labeled training data to realize the capability.
A sensing DNN trained on a design dataset is often observed run-time performance degradations, due to {\em domain shifts} \cite{mathur2018using}. The shifts are generally caused by the deviations of the sensor characteristics and/or the monitored process dynamics of the real deployments from those captured by the design dataset.

Transfer learning \cite{pan2009survey} has received increasing attention for addressing domain shifts. It is a cluster of approaches aiming at storing knowledge learned from one task and applying it to a different but related task. Under the transfer learning scheme, ideally, with little new training data, we can transfer a DNN trained from the {\em source domain} (i.e., the design dataset) to the {\em target domain} (i.e., the sensing data from the real deployment).
However, the prevalent transfer learning techniques, such as {\em freeze-and-train} \cite{guidefortransferlearning} and {\em domain adaptation} \cite{pan2009survey}, require substantial training data collected in the target domain. The freeze-and-train approach
retrains a number of selected layers of a DNN with new target-domain samples to implement the model transfer.
Domain adaptation often needs to train a new DNN to transform the target-domain inference data back to the source domain. For instance, the Mic2Mic \cite{mathur_mic2mic:_2019} trains a cycle-consistent generative adversarial network (CycleGAN) to perform the translation between two microphones that have their own hardware characteristics.
However, the training of CycleGAN requires about 20 minutes of microphone recording from both domains for a keyword spotting task \cite{mathur_mic2mic:_2019}.
In summary, although the prevalent transfer learning techniques reduce the demands on the target-domain training data in comparison with learning from scratch in the target domain, they still need substantial target-domain data to implement the model transfer.

In the cyber-physical sensing applications, both the monitored physical processes and the sensing apparatus are often governed by certain first principles. In this paper, we investigate the approach to exploit such first principles as a form of prior knowledge to reduce the demand on target-domain data for model transfer, vis-{\`a}-vis the aforementioned {\em physics-regardless} transfer learning techniques \cite{guidefortransferlearning, mathur_mic2mic:_2019, pan2009survey}. Recent studies attempt to incorporate prior knowledge in the form of commonsense \cite{yu2017single} or physical laws \cite{stewart2017label,sun2020surrogate} to increase the learning efficiency.
The presentation of the prior knowledge to learning algorithms is the core problem of {\em physics-constrained machine learning}. In \cite{stewart2017label}, the law of free fall is incorporated into the loss function of learning the heights of a tossed pillow in a video. In \cite{sun2020surrogate}, fluid dynamics equations are incorporated into the loss function of training DNNs for real-time fluid flow simulations.
However, these physics-constrained machine learning approaches \cite{stewart2017label,sun2020surrogate} propose new DNN architectures and/or training algorithms; they
are not designed to exploit first principles in transferring existing DNNs to address the domain shift problems.




Nevertheless, the improved learning efficiency of the physics-constrained machine learning
encourages exploiting first principles to address domain shifts more efficiently. To this end, we propose a new approach called
{\em {\bf \em phy}sics-directed data {\bf \em aug}mentation} (PhyAug).
Specifically, we use a minimum amount of data collected from the target domain to estimate the parameters of the first principle governing the domain shift process and then use the parametric first principle to generate augmented target-domain training data. Finally, the augmented target-domain data samples are used to transfer or retrain the source-domain DNN. PhyAug has the following two key features. First, different from the conventional data augmentations that apply unguided {\em ad hoc} perturbations (e.g., noise injection) and transformations (e.g., scaling, rotation, etc) on existing training data to improve the DNNs' robustness against variations, PhyAug augments the training data strategically by following first principles to transfer DNNs. Second, PhyAug uses augmented data to represent the domain shifts and thus requires no modifications to the legacy DNN architectures and training algorithms. This makes PhyAug readily applicable once the data augmentation is completed. In contrast, recently proposed domain adaptation approaches based on adversarial learning \cite{motiian2017few,tzeng2017adversarial,akbari2019transferring,mathur_mic2mic:_2019} update the DNNs under new adversarial training architectures that need extensive hyperparameter optimization and even application-specific redesigns. Such needs largely weaken their readiness, especially when the original DNNs are sophisticated such as the DeepSpeech2 \cite{deepspeech2-impl} for automatic speech recognition.


In this paper, we apply PhyAug to three case studies and quantify the performance gains compared with other transfer learning approaches. The data and code of the case studies can be found in \cite{data-code}. The first and the second case studies aim at adapting DNNs for keyword spotting (KWS) and automatic speech recognition (ASR) respectively to individual deployed microphones. The domain shifts are mainly from the microphone's hardware characteristics.
Our tests show that the microphone can lead to 15\% to 35\% absolute accuracy drops, depending on the microphone quality.
Instead of collecting training data using the target microphone, PhyAug uses a smartphone to play a 5-second white noise and then estimates the frequency response curve of the microphone based on its received noise data.
Then, using the estimated curve, the existing samples in the factory training dataset are transformed into new training data samples, which are used to transfer the DNN to the target domain of the microphone by a retraining process. Experiment results show that PhyAug recovers the microphone-induced accuracy loss by 53\%-72\% and 37\%-70\% in KWS and ASR, respectively.
PhyAug also outperforms the existing approaches including FADA \cite{motiian2017few} that is a domain adaptation approach based on adversarial learning and Mic2Mic \cite{mathur_mic2mic:_2019} and CDA \cite{mathur2018using} that are designed specifically to address microphone heterogeneity. Note that KWS and ASR differ significantly in DNN model depth and complexity.

The third case study is seismic event localization.
In the source domain where the density of the signal propagation medium is spatially homogeneous, the problem of estimating the event location based on the time differences of arrival (TDoAs) of seismic signals received by geographically distributed sensors can follow a multilateration formulation. We aim to adapt to the target domain with an unknown and uneven medium that distorts the TDoAs. Thus, different from the sensor-induced domain shifts in the first and second case studies, the domain shift in this case study is from the variation of the sensed process.
PhyAug estimates the signal propagation slowness model of the medium using a small amount of labeled TDoA data and then generates extensive TDoA data with simulated events to train a DNN for event localization.
Results show that PhyAug only requires 3\% to 8\% of the real labeled TDoA data required by the physics-regardless vanilla approach in achieving the same event localization accuracy.

The main contribution of this paper is the proposed approach of using the first principle fitted with a small amount of source- and target-domain data to extensively augment the target-domain data for model transfer.
This approach is more efficient than the physics-regardless transfer learning in terms of target-domain data sampling complexity.
The applicability of PhyAug is contingent on the availability of the parametric first principle. While the context of cyber-physical systems provides abundant opportunities, the task of pinpointing useful and parametric first principles can be challenging in practice. Fortunately, this task is a one-time effort. Once the first principle for a specific application is identified, the model transfer processes of all the application instances benefit. For instance, by applying PhyAug with a microphone's frequency response curve as the parametric first principle, we can avoid the process of collecting substantial training data from each individual microphone for adapting ASR models.

The remainder of this paper is organized as follows.
\sect\ref{sec:overview} overviews the PhyAug approach and reviews related work. \sect\ref{sec:speech}, \sect\ref{sec:asr}, and \sect\ref{sec:seismic} present the three case studies. \sect\ref{sec:discussion} discusses several issues. \sect\ref{sec:conclude} concludes this paper.

\section{Approach Overview \& Related Work}
\label{sec:overview}

In this section, \sect\ref{subsec:overview} overviews the PhyAug approach. \sect\ref{sec:related} reviews the related studies and explains their relationships with and differences from PhyAug. \sect\ref{subsec:method} discusses the research methodology adopted in this paper.

\subsection{Approach Overview}
\label{subsec:overview}

\begin{figure}[t]
  \includegraphics[width=\columnwidth]{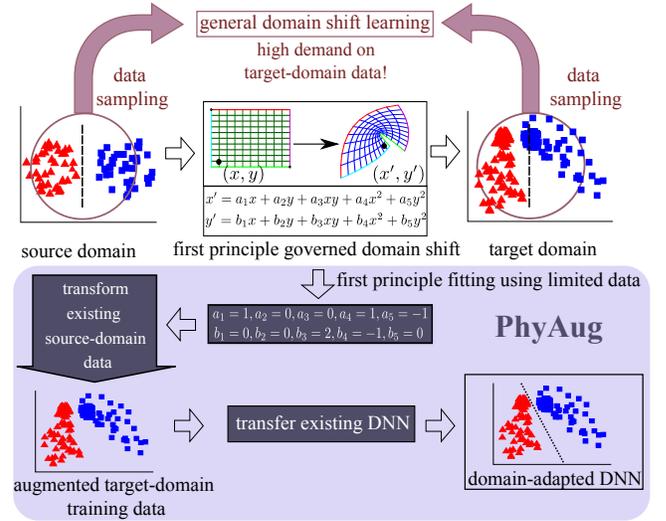}
  \caption{PhyAug workflow.}
  \label{fig:overview}
\end{figure}

\begin{table*}
  \centering
  \caption{Categorization, used techniques, and requirements of various solutions to address domain shifts.}
  \small
  \label{tab:da-compare}
  \begin{tabular}{@{}p{1.3cm}|p{1.8cm}|p{1.6cm}|p{2.3cm}|p{1.5cm}p{1.5cm}p{1.5cm}p{1.2cm}p{2.2cm}@{}}
    \hline
    {\bf Category} & {\bf Used} &  {\bf Solution} & {\bf Applications} & \multicolumn{5}{c}{\bf Requirements} \\
    \cline{5-9}
    & {\bf technique} &  & {\bf in publication} & {\bf Source domain label} & {\bf Target domain label} & {\bf Paired label data} & {\bf First pr- inciple} & {\bf Target-domain data volume}* \\
    \hline
    \center \multirow{7}{*}{\rotatebox[origin=c]{90}{\parbox{2.3cm}{Domain adaptation\\(Model transfer)}}}               & & FADA \cite{motiian2017few} & computer vision & \hfil \ding{52} & \hfil \ding{52} & \hfil \ding{52} & \hfil -- & \ding{121} \ding{121} \ding{121} \ding{121} \\
    \cline{3-9}
    & Adversarial         & ADDA \cite{tzeng2017adversarial} & computer vision & \hfil -- & \hfil -- & \hfil -- & \hfil -- & \ding{121} \ding{121} \ding{121} \ding{121} \ding{121} \ding{121} \ding{121} \ding{121} \ding{121} \ding{121} \\
    \cline{3-9}
    & learning         & TransAct \cite{akbari2019transferring} & activity sensing & \hfil -- & \hfil -- & \hfil -- & \hfil -- & \ding{121} \ding{121} \ding{121} \ding{121} \ding{121} \ding{121} \\
    \cline{3-9}
    &         & Mic2Mic \cite{mathur_mic2mic:_2019} & voice sensing & \hfil -- & \hfil -- & \hfil -- & \hfil -- & \ding{121} \ding{121} \ding{121} \ding{121} \ding{121} \ding{121} \ding{121} \ding{121} \\
    \cline{2-9}
    & Meta learning & MetaSense\cite{gong2019metasense} & voice \& motion & \hfil \ding{52} & \hfil \ding{52} & \hfil -- & \hfil -- & \ding{121} \ding{121} \ding{121} \\
    \cline{2-9}
    & Data & \multirow{2}{*}{\bf PhyAug} & voice sensing & \hfil -- & \hfil -- & \hfil -- & \hfil \ding{52} & \ding{121} \\
    \cline{4-9}
    & augmentation & & event localization & \hfil -- & \hfil \ding{52} & \hfil -- & \hfil \ding{52} & \ding{121} \ding{121}\\
    \hline
    Model ro- & Data & \multirow{2}{*} {CDA \cite{mathur2018using}} & voice and activity & \hfil -- & \hfil -- & \hfil -- & \hfil \ding{52} & \ding{121} \ding{121} \ding{121} \ding{121} \ding{121} \ding{121} \ding{121} \ding{121} \ding{121} \\
    bustness & augmentation &  & sensing &  &  &  &  \\
    \hline
    \multicolumn{9}{l}{$^*$ The bars represent oracle scales partially based on the reported numbers in respective publications. Fully comparable scales are diffi-}\\
    \multicolumn{9}{l}{cult to obtain because the solutions are designed for different applications. PhyAug is compared with FADA, Mic2Mic, and CDA in the}\\

    \multicolumn{9}{l}{evaluation sections of this paper. Reasons for excluding other approaches from the comparison will be discussed in the case studies.}
  \end{tabular}
\end{table*}


Fig.~\ref{fig:overview} illustrates PhyAug's workflow using a simple example, where the DNN performs a two-class classification task based on two-dimensional (2D) data samples and the first principle governing the domain shift is a nonlinear polynomial transform. Such transform can be used to characterize camera lens distortion \cite{pohl2014leveraging}.
To simplify the discussion, this example considers class-independent domain shift, i.e., the transform is identical across all the classes. Note that PhyAug can deal with class-dependent domain shifts, which will be discussed later. As illustrated in the upper part of Fig.~\ref{fig:overview}, the general transfer learning approaches regardless of the first principles need to draw substantial data samples from both the source and target domains. Then, they apply {\em domain shift learning} techniques to update the existing source-domain DNN or construct a prefix DNN \cite{mathur_mic2mic:_2019} to address the domain shift. Extensive data collection in the target domain often incurs undesirable overhead in practice.

Differently, as shown in the lower part of Fig.~\ref{fig:overview}, PhyAug applies the following four steps to avoid extensive data collection in the target domain. \ding{182} The system designer identifies the parametric first principle governing the domain shift.
For the current example, the parametric first principle is
$x' = a_1 x + a_2 y + a_3 xy + a_4 x^2 + a_5 y^2$ and
$y' = b_1 x + b_2 y + b_3 xy + b_4 x^2 + b_5 y^2$,
where $(x, y)$ and $(x', y')$ are a pair of data samples in the source and target domains, respectively, and $a_{i}$, $b_{i}$ are unknown parameters. \ding{183} A small amount of unlabeled data pairs are drawn from the source and target domains. The drawn data pairs are used to estimate the parameters of the first principle.
For this example, if the domain shift is perturbation-free, the minimum number of data pairs needed is the number of unknown parameters of the polynomial transform. If the domain shift is also affected by other unmodeled perturbations, more data pairs can be drawn to improve the accuracy of estimating the parameters under a least squares formulation. If the domain shift is class-dependent, the data pair sampling and parameter estimation should be performed for each class separately. \ding{184} All the existing source-domain training data samples are transformed to the target domain using the fitted first principle, forming an augmented training dataset in the target domain. \ding{185} With the augmented training dataset, various techniques can be employed to transfer the existing DNN built in the source domain to the target domain. For instance, we can retrain the DNN with the augmented data. The retraining can use the existing DNN as the starting point to speed up the process. For instance, for the DeepSpeech2 \cite{deepspeech2-impl} which is a large-scale ASR model used in \sect\ref{sec:asr}, the retraining only requires a half of training time compared with the training from scratch using the augmented data.

For sensing DNN design, the source domain is in general the design dataset. In such case, the source domain cannot be excited any more for data pair sampling in both domains simultaneously. However, we can recreate the excitation to collect the corresponding target-domain samples.
For instance, we can use a speaker to play voice samples in the source-domain dataset and collect the corresponding samples from a target-domain microphone. Similarly, we can use a computer monitor to display image samples in the source-domain dataset and collect the corresponding samples from a target-domain camera that may have optical distortions.




\subsection{Related Work}
\label{sec:related}


The applications of deep learning in embedded sensing systems
have obtained superior inference accuracy compared with heuristics and conventional machine learning.
Various approaches have been proposed to address the domain shift problems in embedded sensing \cite{mathur2018using,mathur_mic2mic:_2019,gong2019metasense,akbari2019transferring} and image recognition \cite{motiian2017few,tzeng2017adversarial}. Table~\ref{tab:da-compare} summarizes the categorization, used techniques, and requirements of these approaches. In what follows, we discuss the important details of these approaches and their differences from PhyAug.

$\blacksquare$ {\bf Domain adaptation:}
Few-shot Adversarial Domain Adaptation (FADA) \cite{motiian2017few} transfers the model with limited amount of target-domain training data. It uses the {\em supervised adversarial learning} technique to find a shared subspace of the data distributions in the source and target domains. FADA requires labeled and paired data samples from both the source and target domains.
Adversarial Discriminative Domain Adaptation (ADDA) \cite{tzeng2017adversarial} uses {\em unsupervised adversarial learning} to learn a feature encoder for the target domain. Although ADDA requires neither class labels nor data pairing, it demands substantial unlabeled target-domain data.
TransAct in \cite{akbari2019transferring} considers sensor heterogeneity in human activity recognition and uses unsupervised adversarial learning to learn stochastic features for both domains. It requires hundreds of unlabeled target-domain data samples.
Mic2Mic \cite{mathur_mic2mic:_2019} applies CycleGAN, which is also an adversarial learning technique, to map the target-domain audio recorded by a microphone ``in the wild'' back to the source-domain microphone for which the DNN is trained.
Mic2Mic requires about 20 minutes of speech recording from both microphones, which represents an overhead. Moreover, it can only perform one-to-one translations. Our experiment results in \sect\ref{sec:speech} and \sect\ref{sec:asr} show that CycleGAN performs unsatisfactorily when the source domains are publicly available speech datasets that are collected using numerous microphones in diverse environments.

PhyAug is a domain adaptation approach. Compared with ADDA \cite{tzeng2017adversarial}, TransAct \cite{akbari2019transferring}, and Mic2Mic \cite{mathur_mic2mic:_2019} that are based on unsupervised adversarial learning and thus require substantial target-domain training data, PhyAug exploits the first principle governing the domain shift to reduce the demand on target-domain data.
Although FADA \cite{motiian2017few} aims at reducing the demand of target-domain data, it requires extensive other information such as class labels in both domains. In contrast, PhyAug can operate with unlabeled data (cf.~\sect\ref{sec:speech} and \sect\ref{sec:asr}).
Different from Mic2Mic \cite{mathur_mic2mic:_2019} that requires the source domain to be a single microphone, PhyAug admits a source-domain dataset collected via many (and even unknown) microphones in the KWS and ASR case studies. This makes PhyAug practical since the datasets used to drive the design of DNNs for real-world applications often consist of recordings from diverse sources.


MetaSense \cite{gong2019metasense} uses data collected from multiple source domains to train a base model that can adapt to a target domain related to the source domains.
However, it requires substantial training data from both domains and class labels from each source domain. For voice sensing, MetaSense cannot use a source-domain dataset collected via many unlabeled microphones. But PhyAug can.

\begin{figure}
  \begin{subfigure}[t]{.47\columnwidth}
    \includegraphics[width=\columnwidth]{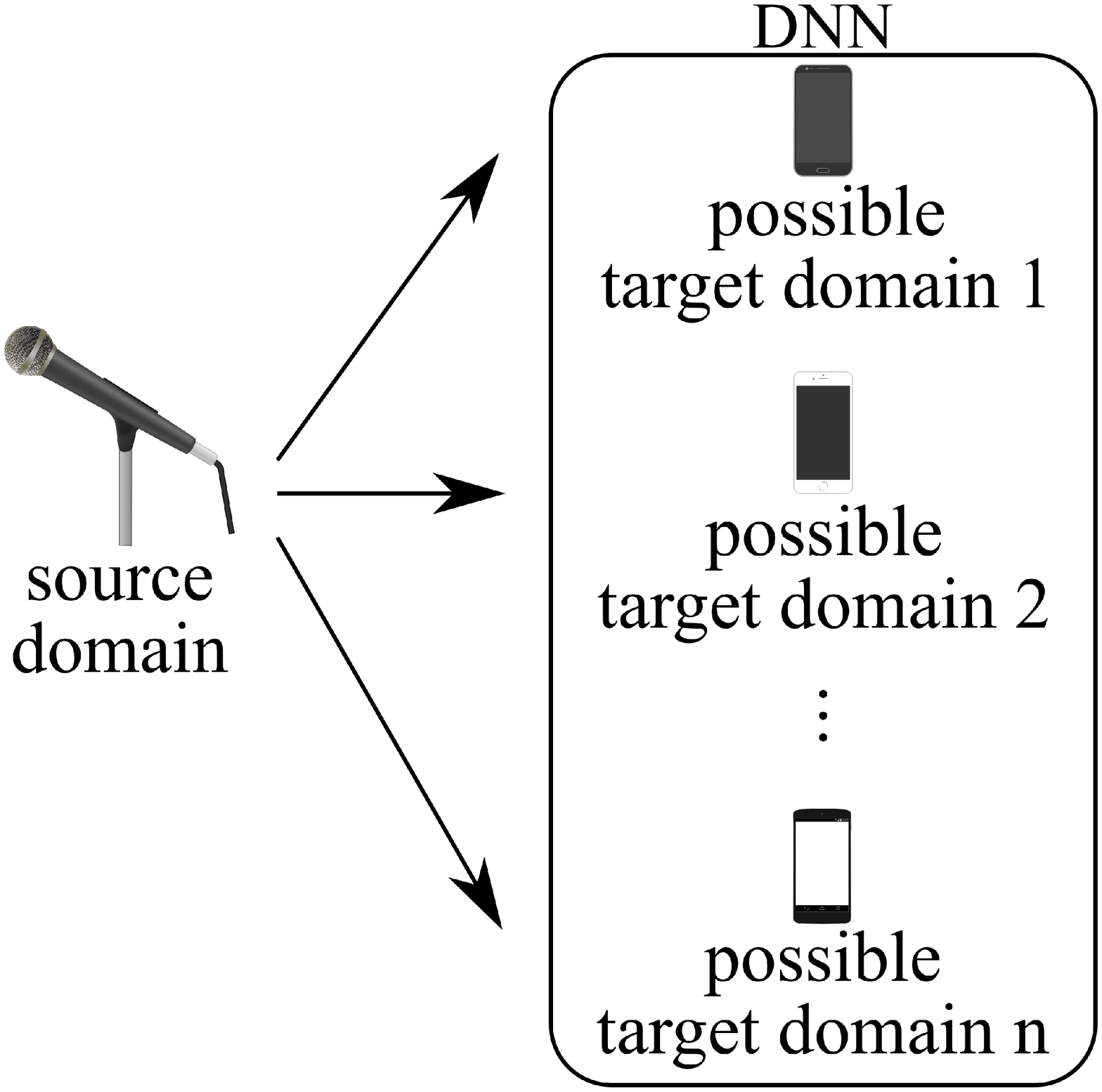}
    \caption{Data augmentation for model robustness (e.g.,\cite{mathur2018using}).}
    \label{fig:conv-da}
  \end{subfigure}
  \hfill
  \begin{subfigure}[t]{.47\columnwidth}
    \includegraphics[width=\columnwidth]{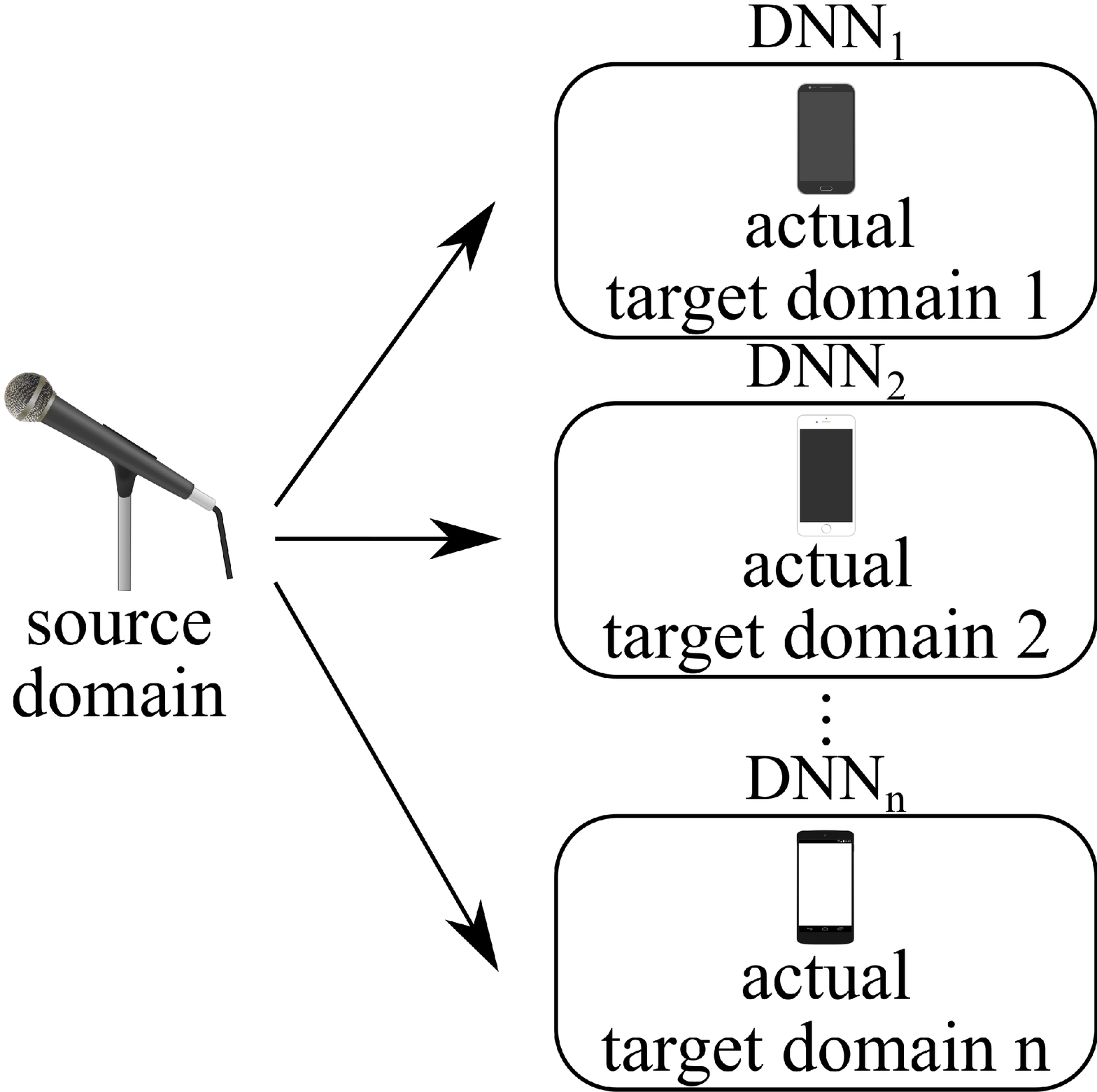}
    \caption{Data augmentation for model transfer (PhyAug).}
    \label{fig:d-da}
  \end{subfigure}
  \caption{Different purposes of data augmentation illustrated using voice sensing. Note that the source domain may contain many microphones used to collect training samples.}
  \label{fig:da-schemes}
\end{figure}

$\blacksquare$ {\bf Model robustness via data augmentation:} Data augmentation has been widely adopted for enhancing model robustness.
As illustrated in Fig.~\ref{fig:conv-da}, a conventional scheme
presumes a number of domain shifts (e.g., scaling, rotation, noise injection, etc) and follows them to generate augmented training samples. Then, the original and the augmented data samples are used to train a single DNN. During the serving phase, this DNN remains robust to the domain shift resembling the presumption. However, should the actual domain shift be out of the presumption, the robustness is lost.
The CDA approach proposed in \cite{mathur2018using} follows the conventional data augmentation scheme to mitigate the impact of sensor heterogeneity on DNN's accuracy. Specifically, it estimates the probability distribution of sensors' heterogeneity characteristics from a {\em heterogeneity dataset} and then uses the characteristics sampled from the estimated distribution to generate augmented training data. As the dataset needs to cover heterogeneity characteristics, its collection in practice incurs a considerable overhead. Specifically, the heterogeneity dataset used in \cite{mathur2018using} consists of 2-hour recordings of 20 different microphones placed equidistant from an audio speaker. If the characteristic of a microphone ``in the wild'' is out of the estimated characteristic distribution (i.e., a missed catch), the enhanced DNN may not perform well. Since CDA uses sensor characteristics, we view it as an approach directed by first principles.

Different from CDA's objective of enhancing model robustness, PhyAug uses data augmentation to transfer a model to a specific target domain. Fig.~\ref{fig:d-da} illustrates this in the context of voice sensing, where microphones' unique characteristics create domains. PhyAug constructs a dedicated DNN for each target domain. Thus, PhyAug is free of the missed catch problem faced by CDA.

\subsection{Methodology}
\label{subsec:method}

As this paper proposes PhyAug which is a domain adaptation approach, it is desirable to show PhyAug's applicability to multiple applications and its scalability to address different levels of pattern sophistication. Therefore, we apply PhyAug to three applications, i.e., KWS, ASR, and seismic event localization. Although KWS and ASR are two specific human voice sensing tasks, they have significantly different complexities. 
Different from KWS and ASR whose domain shift is mainly caused by sensor heterogeneity, the seismic event localization concerns about the domain shift caused by variations of the monitored physical process. For each case study, we also compare PhyAug with multiple existing approaches to show the advantages and performance gains of PhyAug.


\section{Case Study 1: Keyword Spotting}
\label{sec:speech}

Human voice sensing is important for human-computer interactions in many Internet of Things (IoT) applications.
At present, the DNN for a specific human voice sensing task is often trained based on a {\em standard dataset}.
However, as IoT microphones are often of small form factors and low cost, their recordings often suffer degraded and varied voice qualities. In addition, the environment that an IoT microphone resides in can also affect its recording. For instance, the echo patterns in indoor spaces of different sizes can be distinct. Such run-time variations may be poorly captured by the standard dataset. As a result, the DNN yields reduced accuracy after the deployment. We apply PhyAug to address this domain shift problem. Specifically, we start from a swift process of profiling the IoT microphone's frequency response curve (FRC) with the help of a smartphone. Then, we use the FRC to transform the standard dataset. Finally, we retrain the DNN using the transformed dataset to obtain a personalized DNN for the IoT microphone.

In this paper, we consider two human voice sensing functions: KWS and ASR. Most intelligent virtual assistant systems implement both functions. For instance, a virtual assistant often uses a low-power co-processor to perform KWS at all times. Once a designated keyword (e.g., ``Hey Siri'') is detected, the virtual assistant will activate the main processor to execute the more sophisticated ASR. In this section, we focus on KWS. \sect\ref{sec:asr} will focus on ASR. The results show that, with a 5-second smartphone-assisted FRC profiling process, we can recover a significant portion of accuracy loss caused by the domain shifts.

In the case studies of KWS (\sect\ref{sec:speech}) and ASR (\sect\ref{sec:asr}), {\bf source domain} is the standard dataset originally used by the DNN; {\bf target domain} is the dataset of voice samples captured by a specific deployed microphone; {\bf first principle} is the microphone's FRC induced by the microphone hardware and its ambient environment.

\subsection{Problem Description}
\label{subsec:kws-problem}

We conduct a set of preliminary experiments to investigate the impact of diverse microphones on the KWS accuracy. Based on the results, we state the problem that we aim to address.

\begin{figure}
  \centering
  \begin{minipage}[t]{.475\columnwidth}
    \centering
    \vspace{-1.18in}
    \small
    \begin{tabular}{cc|c|c|c|c|c|c|}
      \cline{3-8}
      \rotatebox[origin=c]{90}{$101 \times 40$ MFCC tensor}
      & $\!\!\!\!\!\!\!\Rightarrow\!\!\!\!$
      & \rotatebox[origin=c]{90}{$\;$ 64 $8 \times 20$ conv filters $\;$} & \rotatebox[origin=c]{90}{max pooling (2x2)} & \rotatebox[origin=c]{90}{64 $4 \times 20$ conv filters} & \rotatebox[origin=c]{90}{max pooling (1x1)} & \rotatebox[origin=c]{90}{ 12 ReLUs dense layer } & \rotatebox[origin=c]{90}{softmax}\\
      \cline{3-8}
    \end{tabular}
    \caption{CNN structure used in case study.}
    \label{fig:CNN-keyword}
  \end{minipage}
  \hfill
  \begin{minipage}[t]{.475\columnwidth}
    \centering
    \includegraphics[width=\textwidth]{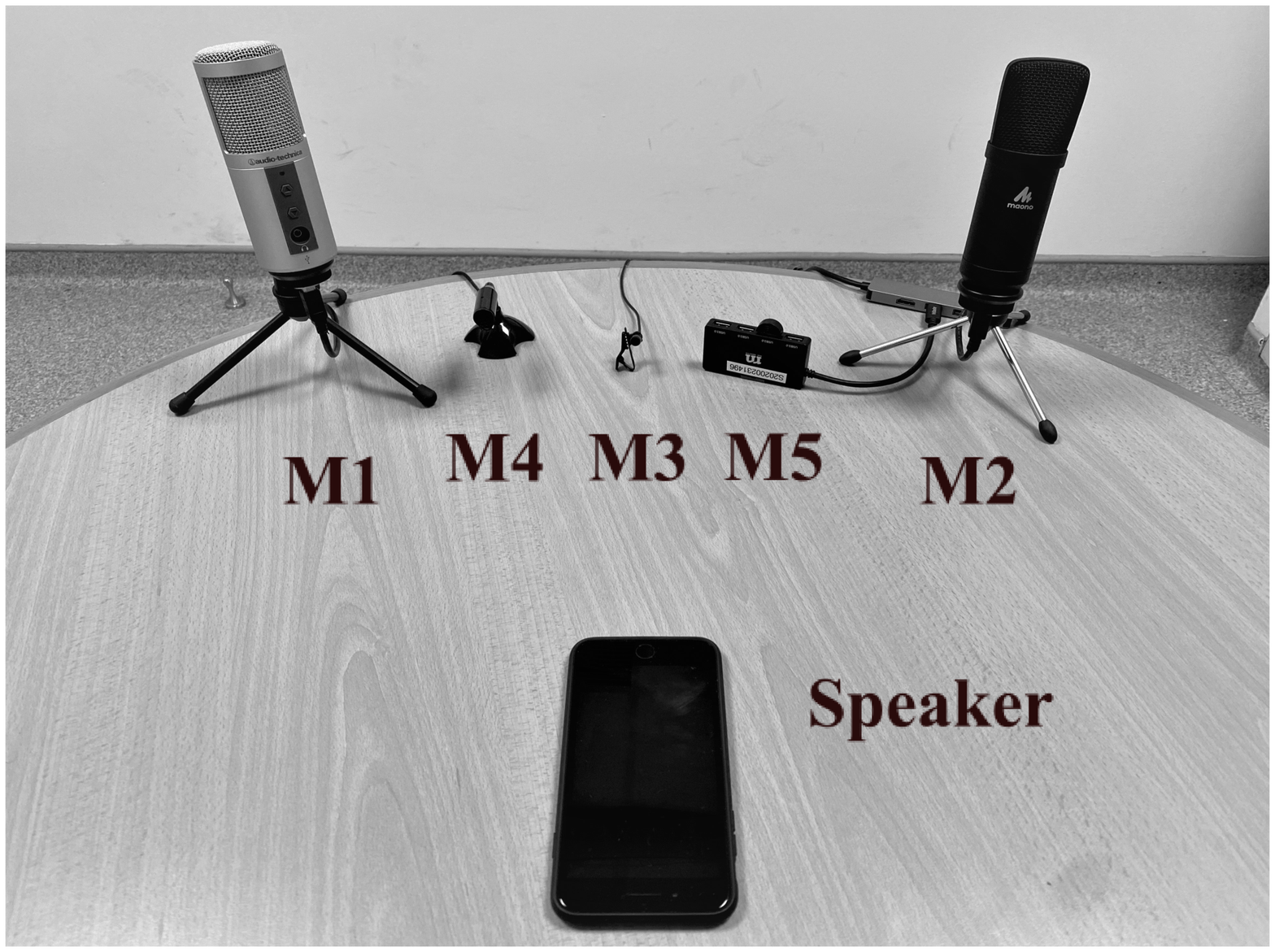}
    \caption{Microphones \& experiment setup.}
    \label{fig:mic-setup}
  \end{minipage}
\end{figure}

\subsubsection{Standard dataset and DNN}
\label{sec:kws-dataset}
We use Google Speech Commands Dataset \cite{warden2018speech} as the standard dataset in this case study. It contains 65,000 one-second utterances of 30 keywords collected from thousands of people. Audio files are sampled at 16 kilo samples per second (ksps). We pre-process the voice samples as follows. First, we apply a low-pass filter (LPF) with a cutoff frequency of $4\,\text{kHz}$ on each voice sample, because human voice's frequency band ranges from approximately $0.3\,\text{kHz}$ to $3.4\,\text{kHz}$. Then, for each filtered voice sample, we generate 40-dimensional Mel-Frequency Cepstral Coefficients (MFCC) frames using 30-millisecond window size and 10-millisecond window shift. The $z$-score normalization is applied on each MFCC frame. Eventually, each voice sample is converted to a $101 \times 40$ MFCC tensor. The dataset is randomly split into training, validation, and testing sets following an 8:1:1 ratio.


We implement a CNN to recognize 10 keywords, i.e., ``yes'', ``no'', ``left'', ``right'', ``up'', ``down'', ``stop'', ``go'', ``on'', and ``off''. We also add two more classes to represent {\em silence} and {\em unknown keyword}.
Fig.~\ref{fig:CNN-keyword} shows the structure of the CNN. It achieves 90\% test accuracy, which is similar to that in \cite{zhang2017hello} and referred to as the {\em oracle test accuracy}.


\begin{figure}[t]
  \begin{minipage}[t]{.475\columnwidth}
    \centering
    \includegraphics[scale=1]{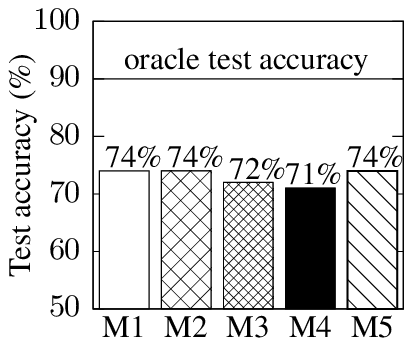}
    \vspace{-0.5em}
    \caption{KWS accuracy on microphones. Horizontal line is accuracy on standard dataset.}
    \label{fig:kws-test-accuracy}
  \end{minipage}
  \hfill
  \begin{minipage}[t]{.475\columnwidth}
    \centering
    \vspace{-1.3in}
    \includegraphics[scale=1.32]{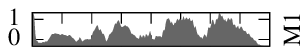}
    \includegraphics[scale=1.32]{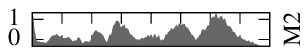}
    \includegraphics[scale=1.32]{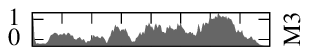}
    \includegraphics[scale=1.32]{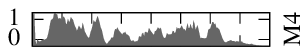}
    \includegraphics[scale=1.32]{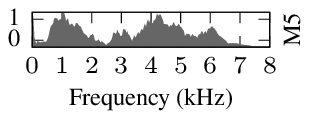}
    \caption{The five microphones' FRCs. The $y$-axis of each sub-figure is normalized amplitude.}
    \label{fig:transfer_func}
  \end{minipage}
\end{figure}

\subsubsection{Impact of microphone on KWS performance}
\label{subsubsec:kws-test}

In this section, we demonstrate that the CNN has performance degradation as a result of microphone heterogeneity.
We test the CNN on samples captured by five different microphones named M1, M2, M3, M4, and M5 as shown in Fig.~\ref{fig:mic-setup} that have list prices from high (\$80) to low (\$3.5). M1 and M2 are two high-end desktop cardioid condenser microphones, supporting sampling rates of $192\,\text{ksps}$ at 24-bit depth and $48\,\text{ksps}$ at 16-bit depth, effective frequency responses of $[30\,\text{Hz}, 16\,\text{kHz}]$ and $[30\,\text{Hz}, 15\,\text{kHz}]$, respectively. M3 is a portable clip-on microphone with an effective frequency response range of $[20\,\text{Hz}, 16\,\text{kHz}]$. M4 and M5 are two low-cost mini microphones without detailed specifications.
Fig.~\ref{fig:mic-setup} shows the placement of the microphones. For fair comparison and result reproducibility, we use an Apple iPhone 7 to play the original samples of the test dataset through its loudspeaker, with all microphones placed at equal distances away.

The samples recorded by each microphone are
fed into the KWS CNN for inference. Fig.~\ref{fig:kws-test-accuracy} shows the test accuracy for each microphone. Compared with the oracle test accuracy of 90\%, there are 14\% to 19\% absolute accuracy drops
due to domain shifts. 
By inspecting the spectrograms of the original test sample and the corresponding ones captured by the microphones, we can observe the differences.
This explains the distinct accuracy drops among microphones.
From the above experiment results, the research questions addressed in this case study are as follows. First, how to profile the characteristics of individual microphones with low overhead? Second, how to exploit the profile of a particular microphone to recover KWS's accuracy?

\subsection{PhyAug for Keyword Spotting}

PhyAug for KWS consists of two procedures: {\em fast microphone profiling} and {\em model transfer via data augmentation}.

\subsubsection{Fast microphone profiling}
\label{subsubsec:fast-profiling}


A microphone can be characterized by its frequency response consisting of magnitude and phase.
We only consider the magnitude component, because the information of a voice signal is largely represented by the energy distribution over frequencies, with little/no impact from the phase of the voice signal in the time domain. Let $X(f)$ and $Y(f)$ denote the frequency-domain representations of the considered microphone's input and output. The FRC to characterize the microphone is $H(f) = \frac{|Y(f)|}{|X(f)|}$, where $|\cdot|$ represents the magnitude.

We propose a fast microphone profiling approach that estimates $H(f)$ in a short time. It can be performed through a factory calibration process or by the user after the microphone is deployed.
Specifically, a loudspeaker placed close to the target microphone emits a band-limited acoustic white noise $n(t)$ for a certain time duration. The frequency band of the white noise generator is set to be the band that we desire to profile. Meanwhile, the target microphone records the received acoustic signal $y_n(t)$. Thus, the FRC is estimated as $H(f) = \frac{|\mathcal{F}[y_n(t)]|}{|\mathcal{F}[n(t)]|}$, where $\mathcal{F}[\cdot]$ represents the Fourier transform. As the white noise $n(t)$ has a nearly constant power spectral density (PSD), this approach profiles the microphone's response at all frequencies in the given band.

In our experiments, we use the iPhone 7 shown in Fig.~\ref{fig:mic-setup} to emit the white noise. We set the frequency band of the noise generator to be $[0, 8\,\text{kHz}]$, which is the Nyquist frequency of the microphone. Fig.~\ref{fig:transfer_func} shows the measured FRCs of the five microphones used in our experiments. Each FRC is normalized to $[0, 1]$. We can see that the microphones exhibit distinct FRCs.
In addition, we observe that the two low-end microphones M4 and M5 have lower sensitivities to the higher frequency band, i.e., $5\,\text{kHz}$ to $8\,\text{kHz}$, compared with the microphones M1, M2, and M3.

\subsubsection{Model transfer via data augmentation}
\label{subsubsec:model-transfer}


We augment training samples in the target microphone's domain by transforming the original training samples using FRC. The procedure for transforming a sample $x(t)$ is as follows: (1) Apply the pre-processing LPF on $x(t)$ to produce $x'(t)$; (2) Conduct short-time Fourier transform using 30-millisecond sliding windows with an offset of 10 milliseconds on $x'(t)$ to produce 101 Fourier frames, i.e., $X_i(f),\,i=1,2,\ldots100$; (3) Multiply the magnitude of each Fourier frame with the FRC to produce $|Y_i(f)| = H(f) \cdot |X_i(f)|$; (4) Generate the MFCC frame from each PSD $|Y_i(f)|^2$; (5) Concatenate all 101 MFCC frames to form the MFCC tensor.
Lastly, PhyAug retrains the CNN with augmented data samples for the microphone. Note that we use the pre-trained CNN as the starting point of the re-training process.

\subsection{Performance Evaluation}
\label{subsec:kws-evaluation}

\subsubsection{Alternative approaches}
\label{subsubsec:kws-baseline}

Our performance evaluation employs the following alternative approaches.

$\blacksquare$ {\bf Data calibration:} At run time, it uses the measured FRC to convert the target-domain data back to the source-domain data and then applies the pre-trained CNN on the converted data. Specifically, let $Y_i(f)$ denote the $i$th Fourier frame after the microphone applies the LPF and short-time Fourier transform on the captured raw data. Then, it estimates the corresponding source-domain PSD as $|X_i(f)|^2 = \left( \frac{|Y_i(f)|}{H(f)} \right)^2$ and generates the MFCC frame from $|X_i(f)|^2$. The MFCC tensor concatenated from the MFCC frames over time is fed to the pre-trained CNN.

$\blacksquare$ {\bf Conventional data augmentation (CDA) \cite{mathur2018using}:} This alternative captures the essence of the approach in \cite{mathur2018using} following the conventional data augmentation scheme illustrated in Fig.~\ref{fig:conv-da}. Specifically, one out of the five microphones, e.g., M1, is designated as the testing microphone. The remaining four, e.g., M2 to M5, are used to generate a {\em heterogeneity dataset} \cite{mathur2018using}. The {\em heterogeneity generator} \cite{mathur2018using} is constructed as follows. 
For each microphone in the heterogeneity dataset, FRC is measured multiple times with the fast profiling process. At any frequency $f$, the FRC value is modeled by a Gaussian distribution. A Gaussian mixture is formed by the four heterogeneity-dataset microphones' Gaussian distributions with equal weights. The Gaussian mixtures for all frequencies form the heterogeneity generator. Then, each source-domain training sample is transformed by an FRC sampled from the heterogeneity generator into an augmented sample. Lastly, the DNN is retrained with the augmented training samples and tested with the samples captured by the testing microphone.

$\blacksquare$ {\bf CycleGAN (essence of \cite{mathur_mic2mic:_2019}):} Mic2Mic \cite{mathur_mic2mic:_2019} trains a CycleGAN using unlabeled and unpaired data samples collected from two microphones $A$ and $B$. Then, CycleGAN can translate a sample captured by $A$ to the domain of $B$, or vice versa. Following \cite{mathur_mic2mic:_2019}, we train a CycleGAN to translate the samples captured by a target microphone to the source domain of Google Speech Commands Dataset. Same as \cite{mathur_mic2mic:_2019}, the training of a CycleGAN for a target microphone uses 15 minutes data collected by the microphone. We train five CycleGANs for the five microphones, respectively. To measure the test accuracy, a test sample collected by a microphone is converted by the corresponding CycleGAN back to the source domain and then fed into the pre-trained CNN.

Compared with PhyAug that requires a single 5-second profiling data collection process for each microphone, CDA repeats the profiling process many times for each heterogeneity microphone to construct the heterogeneity generator; the training of CycleGAN requires 15 minutes of data collected from each target microphone. Thus, both alternative approaches have higher overhead.

  $\blacksquare$ {\bf FADA \cite{motiian2017few}}:
  It trains a feature encoder and classifier in the source domain. Then, it combines source-domain and target-domain data to train a domain-class discriminator. Finally, the weights of the feature encoder and classifier are updated to the target domain through adversarial learning using the domain-class discriminator. To apply FADA for KWS, we follow the architecture in \cite{motiian2017few} and modify the KWS model in Fig.~\ref{fig:CNN-keyword} by adding a fully-connected layer before the last dense layer. Thus, the model has a feature encoder (CNN layers) and a classifier (fully-connected layers).

  We exclude the MetaSense, ADDA and TransAct reviewed in \sect\ref{sec:related} from the baselines for the following reasons. MetaSense cannot be applied to a source-domain dataset collected via many unlabeled microphones. We obtain unsatisfactory results for ADDA in the adversarial training with hours' target-domain training data and extensive hyperparameter tuning. We suspect that the amount of target-domain training data is still insufficient for ADDA. Note that PhyAug only requires five seconds' unlabeled target-domain data as shown shortly. TransAct is customized for activity recognition that differs from human voice sensing.
\subsubsection{Evaluation results}
\label{subsubsec:kws-eval}

\begin{figure}
  \centerline{\includegraphics[scale=1]{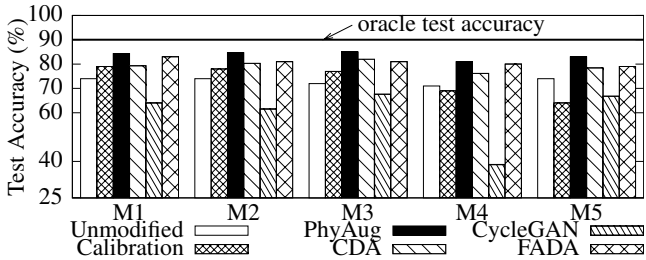}}
  \caption{KWS test accuracy using various approaches on tested microphones. Compared with the unmodified baseline, PhyAug recovers the accuracy losses by 64\%, 67\%, 72\%, 53\%, and 56\% respectively for the five microphones toward the oracle test accuracy.}
  \label{fig:kws-result}
\end{figure}

We apply PhyAug and the alternatives for the five microphones in Fig.~\ref{fig:mic-setup}. The test accuracies are shown in Fig.~\ref{fig:kws-result}. The bars labeled ``unmodified'' are the results from Fig.~\ref{fig:kws-test-accuracy}, for which no domain adaptation technique is applied. We include them as the baseline. The results are explained in detail as follows.

$\blacksquare$ {\bf Data calibration:} It brings test accuracy improvements for M1, M2, and M3. The average test accuracy gain is about 4\%. For the cheap microphones M4 and M5, it results in test accuracy deteriorations. The reason is as follows. Its back mapping uses the reciprocal of the measured FRC (i.e., $1/H(f)$), which contains large elements due to the near-zero elements of $H(f)$. The larger noises produced by the low-end microphones M4 and M5 are further amplified by the large elements of $1/H(f)$, resulting in performance deteriorations. Thus, although this approach may bring performance improvements, it is susceptible to noises.

$\blacksquare$ {\bf PhyAug:} The black bars in Fig.~\ref{fig:kws-result} show PhyAug's results. Compared with the unmodified baseline, PhyAug recovers the test accuracy losses by 64\%, 67\%, 72\%, 53\%, and 56\% for the five microphones. PhyAug cannot fully recover the test accuracy losses. This is because PhyAug only addresses the deterministic distortions due to microphones; it does not address the other stochastic factors such as the environmental noises and the microphones' thermal noises.

$\blacksquare$ {\bf CDA:} It recovers certain test accuracy losses for all microphones. This is because for any target microphone, there is at least one heterogeneity dataset microphone giving a similar FRC as the target microphone. Specifically, from Fig.~\ref{fig:transfer_func}, M1, M2, and M3 exhibit similar FRCs; M4 and M5 exhibit similar FRCs (i.e., they have good responses in lower frequencies).
However, PhyAug consistently outperforms CDA. In addition, CDA introduces larger overhead than PhyAug as discussed in \sect\ref{subsubsec:kws-baseline}.

\begin{figure}
  \centering
  \begin{subfigure}[t]{.5\textwidth}
    \includegraphics[scale=0.43]{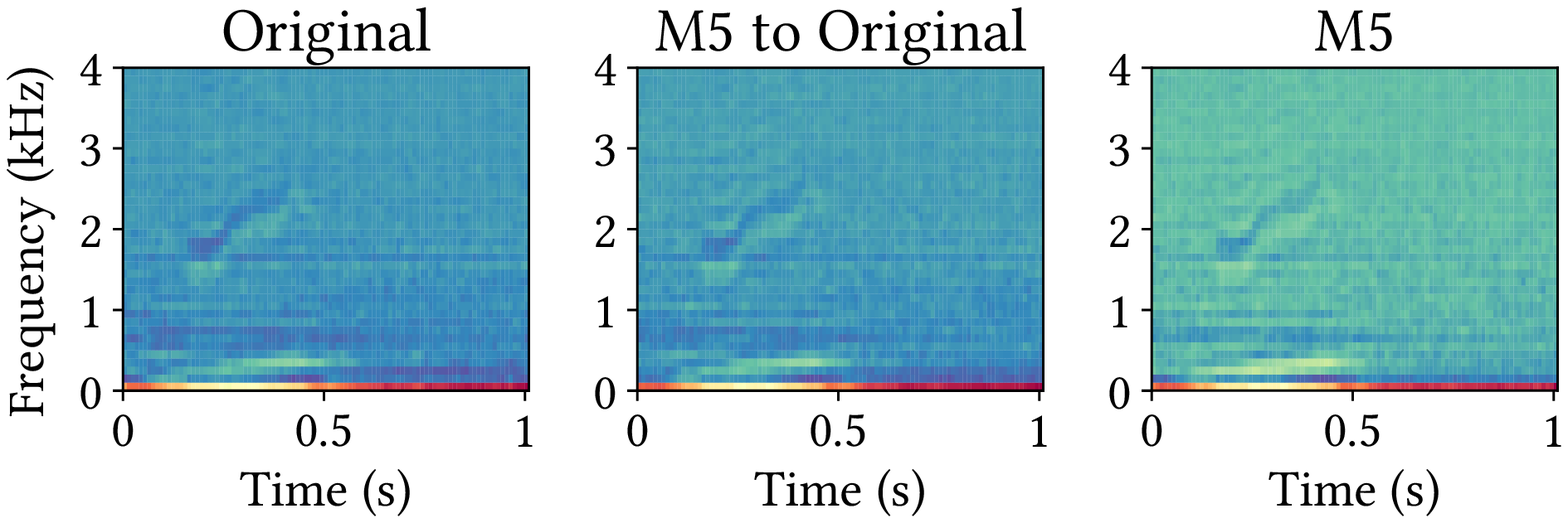}
    \caption{M5 $\longrightarrow$ M1}
    \label{fig:cyclegan-1}
  \end{subfigure}
  
  \vspace{1em}
  \begin{subfigure}[t]{.5\textwidth}
    \includegraphics[scale=0.43]{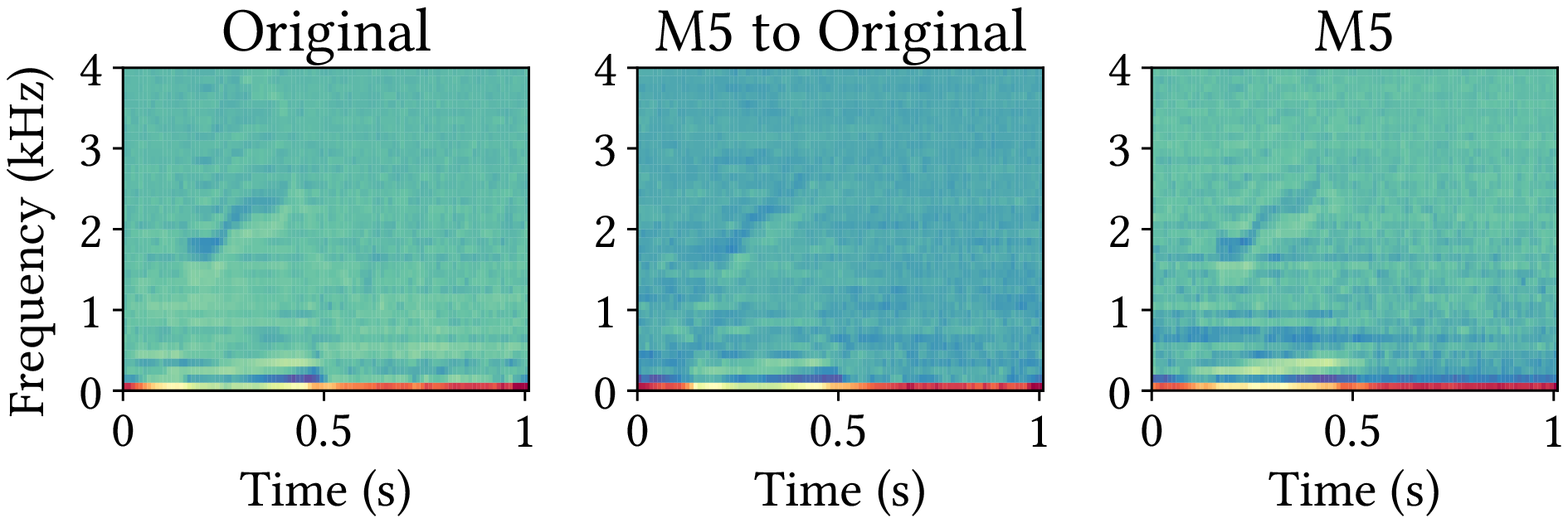}
    \caption{M5 $\longrightarrow$ Original}
    \label{fig:cyclegan-2}
  \end{subfigure}
  \caption{CycleGAN translation results (mid column). (a) Translation from M5 to M1. High similarity between first and second columns shows effectiveness of CycleGAN. (b) Translation from M5 to the domain of Google Speech Commands Dataset. Dissimilarity between first and second columns shows ineffectiveness of CycleGAN.}
\end{figure}

$\blacksquare$ {\bf CycleGAN:} It leads to test accuracy deteriorations for all five target microphones.
Although CycleGAN is effective in translating the domain of a microphone to that of another microphone, which is the basis of Mic2Mic \cite{mathur_mic2mic:_2019}, it is {\em ineffective} in translating a certain microphone to the source domain of a dataset that consists of recordings captured by many microphones. We illustrate this using an example. First, we train a CycleGAN to translate M5 to M1. The first and the third columns of Fig.~\ref{fig:cyclegan-1} shows the spectrograms captured by M1 and M5 for the same sample played by the smartphone in the setup shown in Fig.~\ref{fig:mic-setup}. We can see that there are discernible differences. The mid column shows the output of the CycleGAN, which is very similar to the first column. This result suggests that CycleGAN is effective for device-to-device domain translation and provides a minimal validation of Mic2Mic \cite{mathur_mic2mic:_2019}. Then, we apply the same approach to train a different CycleGAN to translate M5 to the domain of Google Speech Commands Dataset. Fig.~\ref{fig:cyclegan-2} shows the results. The third column is the spectrogram captured by M5 when a dataset sample shown in the first column is played by the smartphone in the setup shown in Fig.~\ref{fig:mic-setup}. The mid column is the CycleGAN's translation result, which has discernible differences from the first column, suggesting the ineffectiveness of CycleGAN. An intuitive explanation is that the CycleGAN shown with samples captured by many microphones during the training phase is confused and caters into no single microphone. Due to the discrepancy between CycleGAN's output and the dataset, the pre-trained CNN fed with CycleGAN's outputs yields low test accuracy.


$\blacksquare$ {\bf FADA:} When we set the number of labeled target-domain samples per class (LTS/C) to 10 for FADA training, it recovers the accuracy loss for the five microphones by 56\%, 38\%, 47\%, 47\%, and 37\%, respectively, as shown in Fig.~\ref{fig:kws-result}.
The performance of FADA increases with LTS/C. When we increase LTS/C to 20, PhyAug still outperforms FADA. Note that PhyAug requires a single unlabeled target-domain sample only.
In addition, from our experience, FADA is sensitive to hyperparameter setting.



$\blacksquare$ {\bf Required noise emission time for microphone profiling:}
In the previous experiments, the microphone profiling uses a 5-minute noise emission time.
We conduct experiments to investigate the impact of shorter noise emission durations on the performance of PhyAug. The results show that the FRCs of a certain microphone measured with various noise emission durations down to five seconds are very similar. The corresponding test accuracies of PhyAug are also similar. (The detailed results are omitted here due to space constraint.) Thus, a noise emission time of five seconds is sufficient. This shows that PhyAug incurs little overhead.

\subsection{Application Considerations}

From the above results, PhyAug is desirable for KWS on virtual assistant systems.
We envisage that more home IoT devices (e.g., smart lights, smart kitchen appliances, etc.) will support KWS.
To apply PhyAug, the appliance manufacturer can offer the fast microphone profiling function as a mobile app and the model transfer function as a cloud service. Thus, the end user can use the mobile app to obtain the FRC, transmit it to the cloud service, and receive the customized KWS DNN. As the KWS DNN is not very deep and PhyAug is a one-time effort for each device, the model retraining performed in the cloud is an acceptable overhead to trade for better KWS accuracy over the entire device lifetime.
  
\section{Case Study 2: Speech Recognition}
\label{sec:asr}

ASR models often have performance degradation after deployments. This section shows the impact of various microphone models on ASR and how PhyAug is applied to recover the accuracy loss.

\subsection{Impact of Microphone on ASR}



We use LibriSpeech \cite{panayotov_librispeech:_2015} as the standard dataset in this case study. It contains approximately 1,000 hours of English speech corpus sampled at $16\,\text{ksps}$. Each sample is an utterance for four to five seconds.
We use an implementation \cite{deepspeech2-impl} of Baidu DeepSpeech2, which is a DNN-based end-to-end ASR system exceeding the accuracy of Amazon Mechanical Turk human workers on several benchmarks. The used DeepSpeech2 model is pre-trained with LibriSpeech training dataset and achieves 8.25\% word error rate (WER) on LibriSpeech test dataset. This 8.25\% WER is referred to as {\em oracle WER}. Note that the input to DeepSpeech2 is the spectrogram of a LibriSpeech sample, which is constructed from the Fourier frames using 20-millisecond window size and 10-millisecond window shift.

DeepSpeech2 has 11 hidden layers with 86.6 million weights. It is far more complicated than the KWS CNN. Specifically, DeepSpeech2 is 175 times larger than the KWS CNN in terms of the weight amount. All the existing studies (e.g., Mic2Mic \cite{mathur_mic2mic:_2019}, MetaSense \cite{gong2019metasense}, and CDA \cite{mathur2018using}) that aimed at addressing domain shift problems in voice sensing only focused on simple tasks like KWS and did not attempt a sophisticated model such as DeepSpeech2.



\begin{figure}[t]
  \centerline{\includegraphics[scale=1.18]{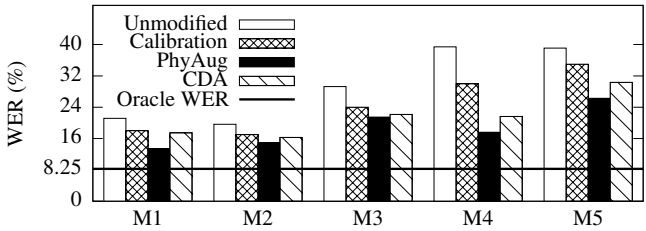}}
  \caption{WERs using various approaches on tested microphones. Compared with the unmodified baseline, PhyAug reduces WER by 60\%, 41\%, 37\%, 70\%, and 42\% respectively for the five microphones toward the oracle WER. As CycleGAN gives very high WERs (about 90\%), it is not shown here.}
  \label{fig:dp2-result}
\end{figure}

We test the performance of the pre-trained DeepSpeech2 on the five microphones M1 to M5 used in \sect\ref{sec:speech}. We follow the same test methodology as presented in \sect\ref{subsubsec:kws-test}.
In Fig.~\ref{fig:dp2-result}, the histograms labeled ``unmodified'' represent the WERs of the pre-trained DeepSpeech2 on the test samples recorded by the five microphones. The horizontal line in the figure represents the {\em oracle WER}. We can see that the microphones introduce about 15\% to 35\% WER increases. In particular, the two low-end microphones M4 and M5 incur the highest WER increases. This result is consistent with the intuition. From the above test results, this section investigates whether PhyAug described in \sect\ref{sec:speech} for KWS is also effective for ASR. Different from the KWS CNN that takes MFCC tensors as the input, DeepSpeech2 takes the spectrograms as the input. Thus, in this case study, PhyAug does not need to convert spectrograms to MFCC tensors in the data augmentation.

\subsection{Performance Evaluation}
\label{sec:asr-eval}
\subsubsection{Comparison with alternative approaches}
We use data calibration, CDA \cite{mathur2018using}, and CycleGAN (i.e., essence of \cite{mathur_mic2mic:_2019}) described in \sect\ref{subsubsec:kws-baseline} as the baselines. FADA \cite{motiian2017few} cannot be readily applied to DeepSpeech2, because FADA requires class labels while DeepSpeech2 performs audio-to-text conversion without the concept of class labels. Differently, PhyAug and the three used baselines transform data without needing class labels.

$\blacksquare$ {\bf Data calibration:} Its results are shown by the histograms labeled ``calibration'' in Fig.~\ref{fig:dp2-result}. Compared with the unmodified baseline, this approach reduces some WERs.


$\blacksquare$ {\bf PhyAug:} Among all tested approaches, PhyAug achieves the lowest WERs for all microphones. Compared with the unmodified baseline, PhyAug reduces WER by 60\%, 41\%, 37\%, 70\%, and 42\%, respectively, for the five microphones toward the oracle WER.

$\blacksquare$ {\bf CDA \cite{mathur2018using}:} It performs better than the data calibration approach but worse than PhyAug. As PhyAug is directed by the target microphone's actual characteristics, it outperforms CDA that is based on the {\em predicted} characteristics that may be inaccurate.

$\blacksquare$ {\bf CycleGAN:} We record a 3.5-hour speech dataset and use it to train a CycleGAN to translate samples captured by a target microphone to the source domain of LibriSpeech dataset. Unfortunately, DeepSpeech2's WERs on the data translated by CycleGAN from the microphones' samples are higher than 90\%, indicating CycleGAN's inefficacy. We are unable to make it effective after extensive attempts. We also try to train the CycleGAN to perform M5-to-M1 domain translation following the design of Mic2Mic in \cite{mathur_mic2mic:_2019}. The resulting WER is 65\%. Although this result is better than 90\%, it is still unsatisfactory.
The reason for CycleGAN's inefficacy for ASR is as follows.
Unlike the KWS task studied in Mic2Mic \cite{mathur_mic2mic:_2019} and \sect\ref{sec:speech} of this paper, which discriminates a few target classes only, end-to-end ASR is much more complicated. CycleGAN may require much more training samples beyond we use to achieve good performance, rendering it too demanding and unattractive in practice.


\subsubsection{Impact of various factors on PhyAug}

We evaluate the impact of the following three factors on PhyAug: the indoor location of the microphone, the distance between the microphone and the sound source, and the environment type. We adopt an evaluation methodology as follows. When we evaluate the impact of a factor, the remaining two factors are fixed. For a certain factor, let $X$ and $Y$ denote two different settings of the factor. We use PhyAug($X$,$Y$) to denote the experiment in which the microphone profiling is performed under the setting $X$ and then the transferred model is tested under the setting $Y$. Thus, PhyAug($X$,$X$) evaluates {\em in situ} performance; PhyAug($X$,$Y$) evaluates the sensitivity to the factor.

\begin{figure}
  \subcaptionbox{Spot A\label{fig:meetroom_spot_A}}{\includegraphics[scale=1]{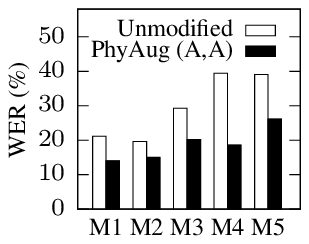}}\hfill%
  \subcaptionbox{Spot B\label{fig:meetroom_spot_B}}{\includegraphics[scale=1]{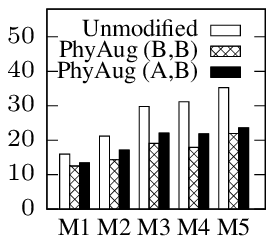}}\hfill%
  \subcaptionbox{Spot C\label{fig:meetroom_spot_C}}{\includegraphics[scale=1]{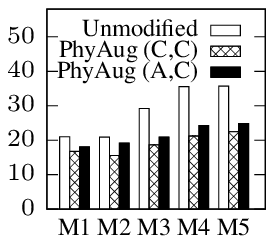}}%
  \caption{PhyAug's {\em in situ} performance and location sensitivity evaluated at three spots in a $7 \times 4\,\text{m}^2$ meeting room.}
  \label{fig:asr-location-impact}
\end{figure}

$\blacksquare$ {\bf Impact of microphone location:} Microphones at different locations of an indoor space may be subject to different acoustic reverberation effects. We set up experiments at three spots, namely, A, B, and C, in a $7 \times 4\,\text{m}^2$ meeting room. Spot B is located at the room center; Spots A and C are located at two sides of B, about $1\,\text{m}$ apart from B along the room's long dimension. The phone and five microphones are set up in the same way as Fig.~\ref{fig:mic-setup}.
Fig.~\ref{fig:asr-location-impact} shows the results of the unmodified baseline approach tested at three spots, as well as PhyAug's {\em in situ} performance and location sensitivity.
PhyAug's {\em in situ} WERs (that is, PhyAug(A,A), PhyAug(B,B), PhyAug(C,C)) are consistently lower than those of the unmodified baseline. The WERs of PhyAug(A,B) and PhyAug(A,C) are slightly higher than PhyAug(B,B) and PhyAug(C,C), respectively. This shows that location affects the performance of a certain ASR model transferred by PhyAug, but not much. Thus, PhyAug for DeepSpeech2 is insensitive to the locations in a certain space.

\begin{figure}
  \subcaptionbox{$75\,\text{cm}$ ($D_1$) \label{fig:meetroom_loc3_15cm}}{\includegraphics[scale=1]{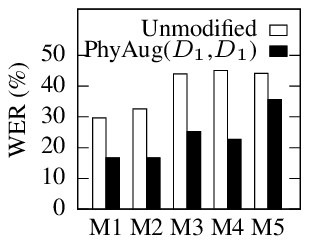}}\hfill%
  \subcaptionbox{$45\,\text{cm}$ ($D_2$) \label{fig:meetroom_loc3_45cm}}{\includegraphics[scale=1]{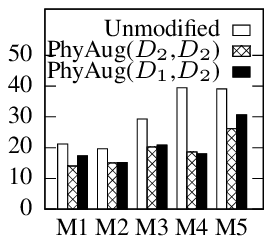}}\hfill%
  \subcaptionbox{$15\,\text{cm}$ ($D_3$) \label{fig:cmeetroom_loc3_75cm}}{\includegraphics[scale=1]{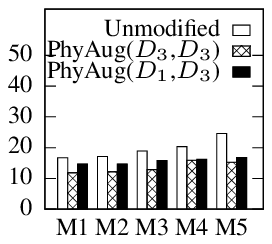}}%
  \caption{PhyAug's {\em in situ} performance and microphone-speaker distance sensitivity evaluated with three distances.}
  \label{fig:asr-distance-impact}
\end{figure}

$\blacksquare$ {\bf Impact of microphone-speaker distance:} The distance affects the signal-to-noise ratio (SNR) received by the microphone and thus ASR performance.
With the setup at the aforementioned Spot C, we vary the distance between the microphones and the iPhone 7 used to play test samples to be $75\,\text{cm}$, $45\,\text{cm}$, and $15\,\text{cm}$ (referred to as $D_1$, $D_2$, and $D_3$).
Fig.~\ref{fig:asr-distance-impact} shows the results.
The unmodified baseline's WERs become lower when the microphone-speaker distance is shorter, due to the increased SNR.
PhyAug's {\em in situ} WERs (i.e., PhyAug($D_1$,$D_1$), PhyAug($D_2$,$D_2$), and PhyAug($D_3$,$D_3$)) are consistently lower than those of the unmodified baseline. The performance gain is better exhibited when the distances are longer. This suggests that {\em in situ} PhyAug improves the resilience of DeepSpeech2 against weak signals.
In most cases, the WERs of PhyAug($D_1$,$D_2$) and PhyAug($D_1$,$D_3$) are slightly higher than those of PhyAug($D_2$,$D_2$) and PhyAug($D_3$,$D_3$), respectively. This shows that the microphone-speaker distance affects the performance of a certain model transferred by PhyAug, but not much. Thus, PhyAug for DeepSpeech2 is insensitive to the microphone-speaker distance.

Another related factor is the speaker's azimuth with respect to the microphone that can affect the quality of the recorded signal due to the microphone's polar-pattern characteristic. For a certain microphone, the different azimuths of the speaker create multiple target domains. If the speaker's azimuth can be sensed (e.g., by a microphone array), PhyAug can be applied. However, as the five microphones used in this paper lacks speaker azimuth sensing capability, we skip the application of PhyAug to address the domain shifts caused by the speaker's azimuth.


\begin{figure}[t]
  \subcaptionbox{Tutorial room\label{fig:TR_loc2_45cm}}{\includegraphics[scale=1]{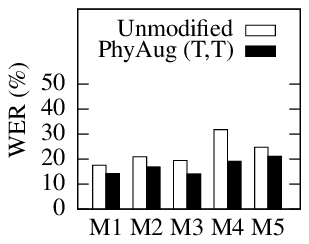}}
  \hfill
  \subcaptionbox{Lecture theater\label{fig:LT_loc2_45cm}}{\includegraphics[scale=1]{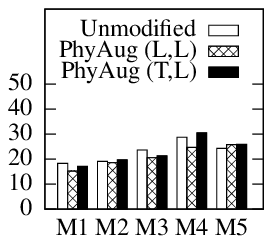}}\hfill
  \subcaptionbox{Open area\label{fig:OA_loc2_45cm}}{\includegraphics[scale=1]{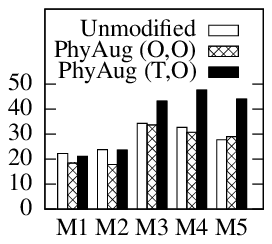}}
  \caption{PhyAug's {\em in situ} performance and environment sensitivity evaluated in three types of environment, namely, small \underline{t}utorial room (T), large \underline{l}ecture theater (L), and outdoor \underline{o}pen area (O).}
  \label{fig:asr-ambient-impact}
\end{figure}

$\blacksquare$ {\bf Impact of environment:} Different types of environments in general have distinct acoustic reverberation profiles, which may affect the microphone's signal reception. We deploy our experiment setup in three distinct types of environments: a small \underline{t}utorial room (T), a large \underline{l}ecture theatre (L), and an outdoor \underline{o}pen area (O).
Fig.~\ref{fig:asr-ambient-impact} shows the results. The unmodified baseline approach has similar results in T and L. Its WERs become higher in O, because O has a higher level of background noise. PhyAug's {\em in situ} WERs in T, i.e., PhyAug(T,T), are consistently lower than those of the unmodified baseline. PhyAug(L,L) and PhyAug(O,O) reduce WERs compared with the unmodified baseline, except for the low-quality microphone M5. As M5 has higher noise levels, the microphone profiling process may not generate fidelity FRCs for M5, leading to increased WERs. As shown in Figs.~\ref{fig:LT_loc2_45cm} and \ref{fig:OA_loc2_45cm}, the WERs of PhyAug(T,L) and PhyAug(T,O) are higher than those of the unmodified baseline. The above results show that PhyAug for DeepSpeech2 may have degraded performance on low-quality microphones. In addition, PhyAug for DeepSpeech2 is sensitive to various environments.

\subsection{Application Considerations}
\label{subsec:impl-considerations}


{\bf Use scenarios:} The results in \sect\ref{sec:asr-eval} show that PhyAug is sensitive to the type of environment because the microphone profiling additionally captures the acoustic reverberation profile of the environment. Thus, PhyAug suits ASR systems deployed at fixed locations, such as residential and in-car voice assistance systems, as well as minutes transcription systems installed in meeting rooms. PhyAug can also be applied to the {\em ad hoc} deployment of ASR and automatic language translation for a multilingual environment.


{\bf PhyAug and continuous learning (CL):} An ASR system can be improved via CL that gradually adapts the ASR model to the speaker and/or the environment when exposed to a continuous data stream for a long period of time. PhyAug is complementary to CL since PhyAug is applied once. Jointly applying PhyAug and CL can maximize the ASR system's quality of service.



\section{Case Study 3: Seismic Source Localization}
\label{sec:seismic}


Estimating the location of a seismic event source using distributed sensors finds applications in earthquake detection \cite{faulkner2011next}, volcano monitoring \cite{werner2006fidelity}, footstep localization \cite{mirshekari2018occupant}, and fall detection \cite{clemente2019indoor}. TDoA-based localization approaches have been widely employed in these applications. The TDoA measurement of a sensor is the difference between the time instants at which the signal from the same event arrives at the sensor and a reference sensor.
In the source domain where the medium density is spatially homogeneous, the seismic signal propagation velocity is also spatially homogeneous. To address measurement noises, the TDoA-based multilateration problem is often solved under a least squares formulation. However, in practice, the medium density is often spatially heterogeneous. This case study aims to deal with the target domain where the medium density is unknown and uneven. For instance, the density of the magma beneath an active volcano varies with depth. As such, seismologists need a {\em slowness model} that depicts the seismic wave propagation speeds at different depths before hypocenter estimation can be performed \cite{lees1991bayesian}. In footstep localization and fall detection, the heterogeneity of the floor materials affects the seismic wave propagation speed and degrades the performance of the simplistic multilateration formulation. Unfortunately, directly measuring the slowness model is tedious or even unfeasible in many cases.

To cope with heterogeneous media, the fingerprinting approach can be employed. Specifically, when a seismic event with a known location is triggered, the TDoA measurements by the sensors form a fingerprint of the known location. With the fingerprints of many locations, a seismic event with an unknown location can be localized by comparing the sensors' TDoA measurements with the fingerprints. The fingerprints can be collected by triggering controlled events at different locations, e.g., controlled explosions in seismology \cite{hupp2016controlled} and hammer excitations in structure health monitoring \cite{hackmann2012holistic}.
Under the fingerprinting approach, determining the location of an event source can be formulated as a classification problem, in which the fingerprint is the input data and the corresponding location is the class label.
To achieve a high localization accuracy, a laborious blanket process of fingerprinting many/all locations is generally required. In this case study, we show that by exploiting the first principle of seismic wave propagation in an uneven medium, we can significantly reduce the amount of fingerprints and achieves a certain level of localization accuracy. Note that, from Appendix~\ref{app:dnn-vs-ls}, even with homogeneous medium, the fingerprinting approach outperforms the least squares approach in terms of response time, while offering comparable localization accuracy.

In this case study, {\bf source domain} is the homogeneous medium for seismic signals; {\bf target domain} is the heterogeneous medium for seismic signals; {\bf first principle} is the slowness model characterizing seismic signal propagations in heterogeneous media.

\subsection{Problem Description}
\label{subsec:seismic-problem}

\begin{figure*}
  \centerline{\includegraphics[scale=1.05]{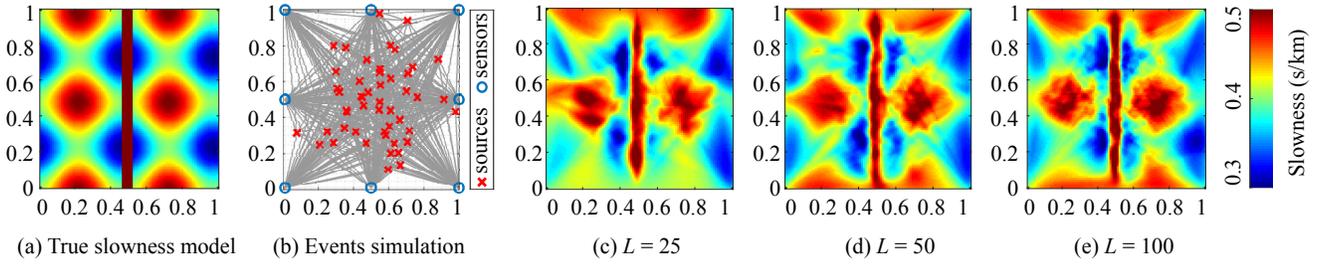}}
  \caption{The $1 \times 1\,\text{km}^2$ 2D field considered in the seismic source localization case study. (a) The ground-truth slowness model with $100 \times 100$ grids. (b) Seismic event source locations and their ray paths to sensors. (c)-(e) The estimated slowness models with 25, 50, and 100 seismic events that occur at random positions in the 2D field as the training samples, respectively.}
  \label{fig:synthetic_model}
\end{figure*}

Consider a 2D field divided into $W_1 \times W_2$ grids, where $W_1$ and $W_2$ are integers. Thus, the field has a total of $N = W_1 \cdot W_2$ grids. Each grid is associated with a slowness value in seconds per kilometer (s/km), which is the reciprocal of the seismic wave propagation speed in the grid. We assume that the slowness at any position in a grid is a constant, while the slownesses in different grids can be distinct. Thus, the slowness model is a matrix (denoted by $\mathbf{S} \in \mathbb{R}^{W_1 \times W_2}$) with the grids' slowness values as the elements. In this case study, we adopt a slowness model from \cite{bianco2018travel} as shown in Fig.~\ref{fig:synthetic_model}(a), which is a $1 \times 1\,\text{km}^2$ square field with a wavy pattern and a barrier stripe in the middle. The pattern and the barrier create challenges to the event localization and will also better exhibit the effectiveness of PhyAug in addressing heterogeneous medium.

There are a total of $M$ seismic sensors deployed in the field. When there is an event occurring in the field, the propagation path of the seismic wave front from the event source to any sensor follows a straight ray path. For instance, Fig.~\ref{fig:synthetic_model}(b) shows the ray paths for the eight sensors considered in this case study. Note that this case study can be also extended to address the refraction of the seismic wave at the boundary of any two grids by using a ray tracing algorithm \cite{lees1991bayesian} to determine the signal propagation path. In Fig.~\ref{fig:synthetic_model}(b), the deployment of the sensors at the field boundary is consistent with the practices of floor event monitoring \cite{mirshekari2018occupant} and volcano activity monitoring \cite{werner2006fidelity}. The seismic event locations
follow a Gaussian distribution centered at the field center.

In what follows, we model the seismic signal propagation process for the $l$th event. For the $m$th sensor, denote the propagation ray path by $p_{l,m}$; denote the {\em ray tracing} matrix by $\mathbf{A}_{l,m} \in \mathbb{R}^{W_1 \times W_2}$, where its $(i,j)$th element is the length of $p_{l,m}$ in the $(i,j)$th grid. If $p_{l,m}$ does not go through the $(i,j)$th grid, the corresponding element of $\mathbf{A}_{l,m}$ is zero. Let $\mathbf{a}_{l,m} \in \mathbb{R}^{1 \times N}$ denote a row vector flattened from $\mathbf{A}_{l,m}$ in a row-wise way. Therefore, the ray tracing matrix for all sensors in the $l$th event, denoted by $\mathbf{A}_l \in \mathbb{R}^{M \times N}$, is $\mathbf{A}_l = [\mathbf{a}_{l,1}; \mathbf{a}_{l,2}; \ldots; \mathbf{a}_{l,M}]$. Let $t_{l,m}$ denote the time for the seismic wave front to propagate from the $l$th event's source to the $m$th sensor. Denote $\mathbf{t}_l = [t_{l,1}; t_{l,2}; \ldots; t_{l,M}] \in \mathbb{R}^{M \times 1}$. Let $\mathbf{s} \in \mathbf{R}^{N \times 1}$ denote a column vector transposed from the row vector that is the row-wise flattening of the slowness model $\mathbf{S}$. Thus, the first principle governing the propagation times is
\begin{equation}
  \mathbf{t}_l = \mathbf{A}_l \mathbf{s}.
  \label{eq:slowness}
\end{equation}

Note that the flattened slowness model $\mathbf{s}$ is identical for all events.
Denote by $\widetilde{\mathbf{t}}_l = [\widetilde{t}_{l,1}; \widetilde{t}_{l,2}; \ldots; \widetilde{t}_{l,M}]$ the measurements of the propagation times. We assume $\widetilde{\mathbf{t}}_l = \mathbf{t}_l + \boldsymbol{\epsilon}$, where the measurement noise $\boldsymbol{\epsilon} \in \mathbb{R}^{M}$ is a random variable following an $M$-dimensional Gaussian distribution $\mathcal{N}\left(\mathbf{0}_M, \sigma_{\epsilon}^{2} \mathbf{I}_M\right)$. In the numerical experiments, we set $\sigma_{\epsilon} = \xi \cdot \bar{\mathbf{t}}_l$, where $\xi$ is called {\em noise level} and $\bar{\mathbf{t}}_l$ is the average value of the elements in $\mathbf{t}_l$. In the evaluation experiments, the default noise level is $\xi = 2\%$.


In the TDoA-based fingerprinting approach, a target-domain training data sample consists of the position of the triggered event as the label and the TDoA measurements as the feature. Specifically, if the first sensor is chosen to be the reference, the feature of the $l$th event is
\begin{equation}
  \mathbf{f}_l = [\widetilde{t}_{l,2} - \widetilde{t}_{l,1}; \widetilde{t}_{l,3} - \widetilde{t}_{l,1}; \ldots; \widetilde{t}_{l,M} - \widetilde{t}_{l,1}] \in \mathbf{R}^{(M-1) \times 1}.
  \label{eq:f}
\end{equation}
A support vector machine (SVM) or DNN can be trained based on a training dataset and then used to localize an event at run time. The research questions addressed in this case study are as follows. First, how to exploit the first principle in Eq.~(\ref{eq:slowness}) to augment the training dataset? Second, to what extent the demand on actual training data samples can be reduced by applying PhyAug?
\subsection{PhyAug for Seismic Source Localization}
\label{subsec:phyaug-seismic}

To use the first principle in Eq.~(\ref{eq:slowness}) to augment the training dataset, the flattened slowness model $\mathbf{s}$ needs to be estimated using some training data samples. This tomography problem can be solved by the Bayesian Algebraic Reconstruction Technique (BART) or Least Squares with QR-factorization (LSQR) algorithm \cite{nocedal2006numerical}. In this work, we apply BART to generate an estimated slowness model denoted by $\hat{\mathbf{s}}$ based on a total of $L$ training samples collected by triggering events with known positions in the field. The details of BART are omitted here due to space constraint and can be found in \cite{rodgers2000inverse}.
Figs.~\ref{fig:synthetic_model}(c)-(e) show $\hat{\mathbf{s}}$ when $L=25$, $L=50$, and $L=100$, respectively.
We can see that when more seismic events are used, the $\hat{\mathbf{s}}$ is closer to the ground truth shown in Fig.~\ref{fig:synthetic_model}(a). The above tomography process uses $L$ labeled target-domain data samples. Thus, PhyAug for this case study requires target-domain class labels as indicated in Table~\ref{tab:da-compare}. As PhyAug can significantly reduce the amount of needed target-domain data samples as shown shortly, the related overhead is largely mitigated.

With the estimated slowness model $\hat{\mathbf{s}}$, we can generate a large amount of augmented fingerprints to extend the training dataset. Specifically, to generate the $x$th augmented fingerprint denoted by $\mathbf{t}_x$, we randomly and uniformly draw a position from the 2D field as the event source location and then compute the ray tracing matrix $\mathbf{A}_x$ and the fingerprint $\mathbf{t}_x = \mathbf{A}_x \hat{\mathbf{s}}$.
Lastly, the SVM or DNN is trained using the extended training dataset consisting of the $L$ genuine training samples and $X$ augmented training samples.

With the above approach, we can generate any number of augmented training samples. In this case study, we adopt the following approach to decide the volume of augmented training samples. Initially, we set $X = 100 \times N$, where $N$ is the number of grids, and train the SVM/DNN with the augmented training dataset. We double the volume of the augmented training samples (i.e., $X = 2 \times X$) until the validation accuracy of the trained SVM/DNN saturates.

\subsection{Performance Evaluation}


We use both SVM and multilayer perceptron (MLP) for finger-print-based source localization. We implement SVM using LIBSVM 3.24 \cite{chang2011libsvm}. It uses radial basis function kernel with two configurable parameters $C$ and $\gamma$.
  During training, we apply grid search to optimize the settings of $C$ and $\gamma$. 
  In addition, we construct a 5-layer MLP. The numbers of neurons in the layers are $M$, 1024, 1024, 512, and $N$, respectively.
  For training, a 0.2 dropout rate is used between any two hidden layers to prevent overfitting. We use cross-entropy as the loss function at the output layer as the training feedback.

\subsubsection{Advantages brought by PhyAug to SVM/MLP-based fingerprinting approach}
\label{subsubsec:phyaug-advantage}

\begin{figure}
  \centering
  \begin{subfigure}[t]{0.475\columnwidth}
      \centering
      \includegraphics{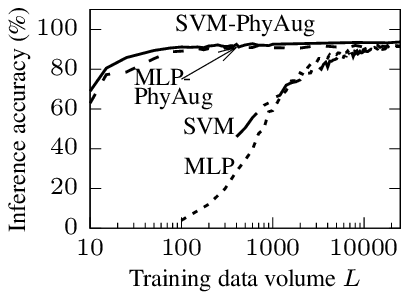}
      \caption{Inference accuracy vs. training data volume for SVM and MLP with or without PhyAug.}
      \label{fig:svm-mlp-eval-2pct}
  \end{subfigure}%
  \hfill
  \begin{subfigure}[t]{.475\columnwidth}
    \includegraphics{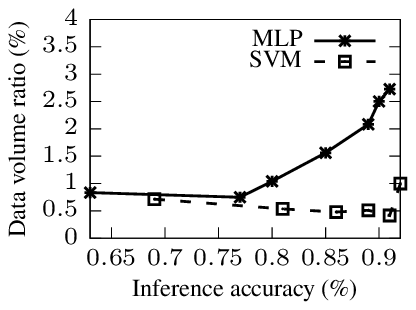}
    \caption{Ratio of training data volumes with and without PhyAug vs. required inference accuracy.}
    \label{fig:training-data-compare-2pct}
  \end{subfigure}
  \caption{Impact of PhyAug on SVM/MLP-based fingerprinting approaches (number of grids: 400; $\xi=2\%$).}
\end{figure}

We set $N = 20 \times 20$. We use the grid-wise inference accuracy as the evaluation metric. Fig.~\ref{fig:svm-mlp-eval-2pct} shows the inference accuracy of SVM and MLP, without and with PhyAug, versus the training data volume $L$. First, we discuss the results of SVM and MLP without PhyAug. We can see that, the inference accuracy of SVM and MLP increases with $L$. When more than 8,000 training samples are provided, SVM and MLP achieve more than 92\% inference accuracy. When less than 11,000 training samples are provided, SVM outperforms MLP; otherwise, MLP outperforms SVM. This observation is consistent with the general understanding that deep learning with ``big data'' outperforms the traditional machine learning approaches. Second, we discuss the results of SVM and MLP with PhyAug. The inference accuracy of SVM-PhyAug and MLP-PhyAug also increases with $L$. With more training samples, the estimated slowness model $\hat{\mathbf{s}}$ is more accurate. As a result, the augmented data samples will be of higher quality, thus helping the SVM and MLP achieve higher test accuracy. From Fig.~\ref{fig:svm-mlp-eval-2pct}, we can see that PhyAug boosts the inference accuracy of SVM and MLP when the training data volume is limited. Fig.~\ref{fig:training-data-compare-2pct} shows the ratio of the training data volumes required by a classifier with or without PhyAug to achieve a specified inference accuracy. With PhyAug, only less than 3\% training samples are needed. This shows that PhyAug is very effective in reducing the demand on training data.

\subsubsection{Impact of noise level}

The noise level of TDoA data affects the accuracy of $\hat{\mathbf{s}}$. Our evaluation results in Appendix~\ref{sec:noise-impact} show that PhyAug requires 1\% to 8\% of actual training data required by SVM or MLP when $\xi$ increases from 0 to 8\%.

\subsubsection{Summary}

Different from the KWS and ASR case studies that use PhyAug to recover recognition accuracy loss mainly caused by sensor hardware characteristics, this case study uses PhyAug to reduce the demand for actual training data in dealing with the complexity of the sensed physical process. Although this case study is primarily based on numerical experiments, the results provide baseline understanding on the advantages brought by PhyAug.



\section{Discussions}
\label{sec:discussion}


The three case studies have demonstrated the advantages of exploiting the first principles in dealing with domain shifts that are often experienced by deployed sensing systems. Pinpointing the useful first principles can be challenging in practice and requires separate studies/experimentation for different applications. For the applications that lack useful first principles, we may fall back to the existing physics-regardless transfer learning approaches. However, the fallback option should not discourage us from being discerning on the exploitable first principles in the pursuit of advancing and customizing deep learning-based sensing in the domain of physics-rich cyber-physical systems. In what follows, we briefly mention several other sensing tasks that PhyAug may be applicable to, which are also interesting for future work.

$\blacksquare$ Polynomial transforms can describe the optical distortions of the camera lens that may be introduced purposely (e.g., fisheye lens) \cite{pohl2014leveraging}. Visual sensing applications can adapt to varied optical distortions to improve DNN performance.

$\blacksquare$ Room impulse response (RIR) describes indoor audio processes. Smart voice-based appliances can exploit RIR as the first principle for effective adaptations to the deployment environments. Acoustic-based indoor localization with deep learning \cite{song2018deep} can exploit RIR to reduce target-domain training data sampling complexity.

$\blacksquare$ Computational fluid dynamics (CFD) describes the thermal processes in indoor spaces (e.g., data centers). A trained deep reinforcement learning-based environment condition controller can adapt to new spaces with CFD models and a few target-domain data samples in each new space.

\section{Conclusion}
\label{sec:conclude}

This paper described PhyAug, an efficient data augmentation approach to deal with domain shifts governed by first principles. We presented the applications of PhyAug to three case studies of keyword spotting, automatic speech recognition, and seismic event localization. They have distinct task objectives and require deep models with quite different architectures and scales. The extensive and comparative experiments showed that PhyAug can recover significant portions of accuracy losses caused by sensors' characteristics and reduce target-domain training data sampling complexity in dealing with the domain shifts caused by the variations of the dynamics of the sensed physical process.


\begin{acks}
  This research was conducted at Singtel Cognitive and Artificial Intelligence Lab for Enterprises (SCALE@NTU), which is a collaboration between Singapore Telecommunications Limited (Singtel) and Nanyang Technological University (NTU) that is funded by the Singapore Government through the Industry Alignment Fund - Industry Collaboration Projects Grant.
\end{acks}


\bibliography{sample-base}
\bibliographystyle{ACM-Reference-Format}

\appendix
\section{SVM/DNN vs. Least Squares Method for Seismic Source Localization}
\label{app:dnn-vs-ls}

\begin{figure}[h]
  \centering
  \begin{subfigure}[t]{.475\columnwidth}
    \includegraphics{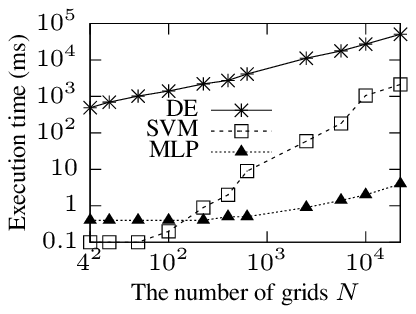}
    \caption{Inference time vs. the number of grids (both the x- and y-axes are in log scale).}
    \label{fig:de-model-execution-time}
  \end{subfigure}
  \hfill
  \begin{subfigure}[t]{.475\columnwidth}
    \includegraphics{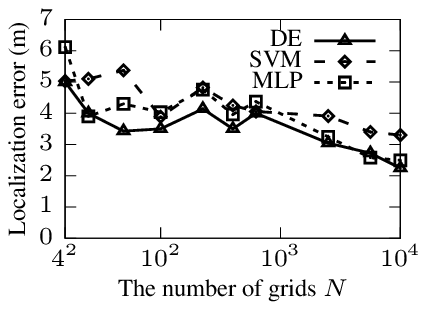}
    \caption{Localization error vs. the number of grids (the x-axis is in log scale).}
    \label{fig:algo-localization-error}
  \end{subfigure}
  \caption{Performance comparison of differential evolution (DE), SVM, and MLP.}
  \label{fig:performance-compare}
\end{figure}

The estimated slowness model $\hat{\mathbf{s}}$ can be directly used to estimate the source location at run time by a least squares method. In the least squares method, 
we apply differential evolution (DE), which is a population-based metaheuristic search algorithm, to perform grid-granular search and iteratively improve a candidate solution $p$ with $\| \mathbf{f} - f\left(\mathbf{A}_{p} \hat{\mathbf{s}} \right) \|_{\ell_2}^2$ as the error metric. In the above error metric, the $\mathbf{f}$ is the feature vector of TDoA measurements given by Eq.~(\ref{eq:f}) for the run-time event; the $\mathbf{A}_p$ is the ray tracing matrix of the candidate position $p$; the $f(\cdot)$ is a function converting the seismic propagation times to the feature vector of TDoA measurements. 
Fig.~\ref{fig:performance-compare} compares the performance of DE, SVM, and MLP in terms of average execution time over 100 events.
In the evaluation, we increase the number of grids $N$ for finer inference granularity.
Fig.\ref{fig:de-model-execution-time} shows the execution time versus $N$.
From a regression analysis on the results, DE has a time complexity of $O(N^{0.45})$.
It's execution time is several orders of SVM and MLP.
For instance, when $N = 22500$, DE's execution time is $50.73\,\text{s}$, which is about 23x and 12,500x longer than SVM's and MLP's, respectively. The long response delays make DE unsuitable for a range of time-critical applications such as earthquake early warning \cite{faulkner2011next}.
\balance
From Fig.~\ref{fig:de-model-execution-time}, the execution time of ML is within $10\,\text{ms}$ when $N$ is up to 22,500.
Fig.~\ref{fig:algo-localization-error} shows the average localization error in terms of Euclidean distance versus $N$.
We can see that the three approaches give comparable localization accuracy. 
From the above results, SVM and MLP are superior to DE due primarily to response times.

\section{Impact of Noise Level on PhyAug for Seismic Source Localization}
\label{sec:noise-impact}

\begin{figure}[t]
  \centering
  \begin{subfigure}[t]{0.49\columnwidth}
    \includegraphics{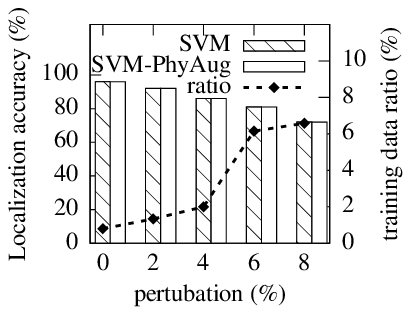}
    \caption{Classifier is SVM.}
    \label{fig:svm-perturb-eval}
  \end{subfigure}
  \hfill
  \begin{subfigure}[t]{0.49\columnwidth}
      \centering
      \includegraphics{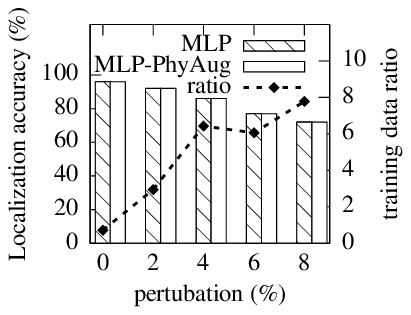}
      \caption{Classifier is MLP.}
      \label{fig:MLP-perturb-eval}
    \end{subfigure}
  \caption{Impact of TDoA measurement noise level on PhyAug's effectiveness.}
  \label{fig:perturb-eval}
\end{figure}

This appendix contains experiment results on the impact of noise level $\xi$ on the performance of PhyAug for seismic source localization. As defined in \sect\ref{subsec:seismic-problem}, the TDoA measurement contains a random noise following $\mathcal{N}\left(\mathbf{0}_M, (\xi \bar{\mathbf{t}}_l)^2 \mathbf{I}_M\right)$.
The histograms in Fig.~\ref{fig:perturb-eval} show the grid-wise localization accuracy of SVM, MLP, and their PhyAug-assisted variants when $\xi$ increases from 0\% to 8\%. The dashed curve in Figs.~\ref{fig:svm-perturb-eval} and \ref{fig:MLP-perturb-eval} shows the ratio between the volumes of actual training data required by SVM/MLP with and without PhyAug. The SVM/MLP approach uses the same amount of training data for all $\xi$ settings, whereas we adjust the amount of the actual training data used for the PhyAug-assisted variant to achieve the same grid-wise localization accuracy as the SVM/MLP approach. From the figure, the localization accuracy decreases with $\xi$. This is consistent with intuition because larger noise levels lead to more classification errors. In addition, the ratio of the actual training data amounts required by SVM/MLP with and without PhyAug increases with $\xi$.
For example,
MLP-PhyAug only requires about 1\% of the training data needed by MLP without PhyAug to achieve the same 96\% accuracy when $\xi = 0\%$; this ratio increases to about 8\% to achieve the same 77\% accuracy when $\xi = 8\%$.
This is because PhyAug needs more actual training data to estimate a good slowness model when the noise level is higher.
Nevertheless, PhyAug reduces the demand for actual training data by a factor of more than 10 when $\xi$ is up to 8\%.

\end{document}